\documentclass[a4paper,fleqn,usenatbib]{mnras}

\usepackage[T1]{fontenc}
\usepackage{ae,aecompl}
\usepackage{color}

\usepackage{graphicx}	
\usepackage{amsmath}	
\usepackage{amssymb}	

\usepackage{newtxtext,newtxmath}

\title[Analytical Wind Solutions II: The $\delta$-slow Regime]{Analytical Solutions for Radiation-Driven Winds in Massive Stars II: The $\delta$-slow Regime}

\author[I. Araya et al.]{
I. Araya$^{1}$\thanks{E-mail: ignacio.araya@umayor.cl},
A. Christen$^{2}$,
M. Cur\'e$^{3}$,
L. S. Cidale$^{4, 5}$,
R. O. J. Venero$^{4, 5}$,
C. Arcos$^{3}$,
\newauthor
A. C. Gormaz-Matamala$^{3,7}$,
M. Haucke$^{4, 5}$,
P. Esc\'arate $^{8}$ ,
and H. Claver\'ia$^{9}$ 
\\
$^{1}$ Centro de Investigaci\'on DAiTA Lab, Facultad de Estudios Interdisciplinarios, Universidad Mayor, Chile\\
$^{2}$ Instituto de Estad\'istica, Facultad de Ciencias, Universidad de Valpara\'{\i}so,
Av. Gran Breta\~na 1111, Valpara\'{\i}so, Chile\\
$^{3}$ Instituto de F\'{\i}sica y Astronom\'{\i}a, Facultad de Ciencias, Universidad de Valpara\'{\i}so,
Av. Gran Breta\~na 1111, Valpara\'{\i}so, Chile\\
$^{4}$ Departamento de Espectroscop\'{\i}a, Facultad de Ciencias Astron\'omicas y Geof\'{\i}sicas, Universidad Nacional de La Plata (UNLP), Argentina\\
$^{5}$ Instituto de Astrof\'{\i}sica La Plata, CCT La Plata, CONICET-UNLP, La Plata, Argentina\\
$^{7}$ Departamento de Ciencias, Facultad de Artes Liberales, Universidad Adolfo Ib\'{a}\~{n}ez, Av. Padre Hurtado 750, Viña del Mar, Chile\\
$^{8}$ Instituto de Electricidad y Electr\'onica, Facultad de Ciencias de la Ingenier\'ia, Universidad Austral de Chile, General Lagos 2086, Valdivia, Chile\\
$^{9}$ Instituto de Estad\'{i}stica, Pontificia Universidad Cat\'{o}lica de Valpara\'{i}so, Chile
}
\date{Accepted XXX. Received YYY; in original form ZZZ}

\pubyear{2021}

\begin{document}
\label{firstpage}
\pagerange{\pageref{firstpage}--\pageref{lastpage}}
\maketitle

\begin{abstract}
	Accurate mass-loss rates and terminal velocities from massive stars winds are essential to obtain synthetic spectra from radiative transfer calculations and to determine the evolutionary path of massive stars. From a theoretical point of view,  analytical expressions for the wind parameters and velocity profile would have many advantages over numerical calculations that solve the complex non-linear set of hydrodynamic equations.  In a previous work, we obtained an analytical description for the fast wind regime. Now,  we propose an approximate expression for the line-force in terms of new parameters and obtain a velocity profile closed-form solution (in terms of the Lambert $W$ function) for  the $\delta$-slow regime. Using this analytical velocity profile, we were able to obtain the mass-loss rates based on the m-CAK theory. Moreover, we established a relation between this new set of line-force parameters with the known stellar and m-CAK line-force parameters. To this purpose, we calculated a grid of numerical hydrodynamical models and performed a multivariate multiple regression. The numerical and our descriptions lead to good agreement between their values.  
\end{abstract}

\begin{keywords}
hydrodynamics --- methods: analytical --- stars: early-type --- stars: mass-loss --- stars: winds, outflows
\end{keywords}

\section{Introduction}

The knowledge of stellar wind properties of massive stars is fundamental for understanding stellar evolution processes, different evolutionary scenarios and enrichment of star's  nearby environments. 

Accurate wind parameters (mass-loss rate and terminal velocity) are crucial for the study of the wind properties of massive stars. Insights into the physics of stellar winds are attained by studying the effects of wind parameters on the emergent line spectrum and by comparing the latter with observations. From a theoretical point of view, this implies to solve highly non-linear equations in which the radiation field and hydrodynamics are strongly coupled. 

Winds of massive stars are driven by the transfer of momentum from the radiation field to the plasma by scattering processes in the spectral lines \citep{lucy1970}. Currently, these winds  are best described by the  m-CAK theory \citep{castor1975,friend1986,ppk1986}. 

Generally, there are many approximations that reduce considerably the complexity of the computation of the hydrodynamic  and the NLTE radiative transfer solutions. One example is the extensive use of a simple analytical approximation for the velocity field, the so-called $\beta$-law, first proposed by \citet{lamers1978}. A value of $\beta \simeq 0.8 - 1.2$, generally agrees very well with the m-CAK numerical hydrodynamic solution \citep{lamers1999}. This value of $\beta$ is determined empirically by fitting the observed line profile with a synthetic one. This approximation has been proved to be very effective and efficient to describe the winds of O- and early B-type supergiants. However, in the case of late B- and A-type supergiants there is a clear tendency towards higher values of $\beta$, even with values larger than $3$, leading to inconsistencies with respect to the hydrodynamic theory \citep{stahl1991,verdugo1999,crowther2006,lefever2007,markova2008,searle2008,haucke2018}. Therefore, accurate analytical approximations of the m-CAK hydrodynamic equations are indispensable  to have  a self-consistent coupling between the hydrodynamics and multidimensional radiative transfer problems in moving media. 

For the case of the fast regime (standard m-CAK solution), this issue was addressed by \citet{villata1992}, \citet{muller2008} and \citet{araya2014}. The aim of this work is to extend the procedure of \citet{araya2014} to the $\delta$-slow regime. The $\delta$-slow solution \footnote{Previously, \citet{cure2004} found another type of slow solution for rapidly rotating stars, called $\Omega$-slow solution.}, found by \citet{cure2011}, is based on the m-CAK  theory, that describes the wind velocity profile when the ionization-related line-force parameter $\delta$ takes higher values than the ones provided by the standard m-CAK solution \citep[see, e.g.,][and references therein]{lamers1999}. High values of $\delta$, even larger than $1/3$, which corresponds to a wind with neutral hydrogen as a trace element \citep{puls2000}, are expected in strong ionization gradients \citep[see also][]{kudritzki2002}. The $\delta$-slow solution is characterized by low terminal speeds ($v_{\infty}$) and might explain the obtained values for late-B and A-type supergiants. This solution also seems to fit quite well the observed anomalous correlation between the terminal and escape velocities  found in A supergiants, as well as their corresponding  wind momentum-luminosity relationship \citep{cure2011}. 

With the purpose to have an approximate solution from the hydrodynamic, \citet{araya2014} developed an expression in terms of the stellar and m-CAK line-force parameters ($\alpha$, $k$, and $\delta$) and applied it to the fast regime. This expression, based on the   works of \citet{muller2008} and \citet{villata1992}, describes the line acceleration as function of the radial distance, allowing  to solve analytically the hydrodynamic stationary equation of motion. The use of  expressions for both radiation force and velocity profile as a function of the line-force parameters can provide a clear view into how the line-driven mechanism is related with the hydrodynamics. 

On the other hand, it is important to obtain a simple representation of the radiation force and the derived slow solutions under such different ionization conditions. Therefore, a significant contribution of this work consists in offering a quick way to generate an analytical expression to estimate mass-loss rates for these alternative wind regimes. There are currently no parametric expressions that can be used for this purpose without the need to fully solve the hydrodynamic equations.

This work is organized as follows: Section \ref{Sbasic-eq} presents briefly the hydrodynamic equations for line-driven winds and the dimensionless form of the equation of motion. In Section \ref{SSgline}, the basic concepts developed by \citet{muller2008} are recapitulated including their line acceleration term as function of the radial distance. Then, this line acceleration term is modified with the purpose to obtain a better agreement with the $\delta$-slow solution. In Section \ref{LAP}, a recipe to obtain the line acceleration parameters (required by the line acceleration term) is developed, based on a grid of hydrodynamic models and a multivariate multiple regression. Then, an analytical expression for the $\delta$-slow solution is developed and compared with the numerical models described in Section \ref{Sresults}.  In Section \ref{Sconclu}, we give our conclusions.  In addition,  a recipe to derive the mass-loss rate based on our expression is provided in Appendix \ref{mdotcure}.

\section{The Standard Hydrodynamical Wind Model}
\label{Sbasic-eq}
The CAK theory for line-driven winds was originally developed by \citet{castor1975}. This theory describes, for a point source, a stationary, one-dimensional, non-rotating, isothermal, outflowing wind with spherical symmetry. Adopting these assumptions, and neglecting the effects of viscosity, heat conduction and  magnetic fields, the equations of mass conservation and radial momentum state:

\begin{equation}
\label{Scontinuity}
4\, \pi \, r^{2}\, \rho \, v = \dot{M}, 
\end{equation}

\noindent and

\begin{equation}
\label{Smomentum}
v \, \frac{dv}{dr}=-\frac{1}{\rho}\frac{dp}{dr} - \frac{G\, M_{*} (1-\Gamma_{\rm{E}})}{r^{2}} + g^{\rm{line}}.
\end{equation}

\noindent Here $v$ is the fluid radial velocity, $dv/dr=v'$ is the velocity gradient and $g^{\rm{line}}$ is the line acceleration. All other variables have their standard meaning \citep[for a detailed derivation and definitions of variables, constants and functions, see][]{cure2004}. 

The so called m-CAK theory, which include the effects of rotation and a disk-like source, was developed by  \citet{friend1986} and \citet{ppk1986}, based on a general expression from \citet{abbott1982} for the line force: 
\begin{equation}
g^{\rm{line}} = \frac{C}{r^{2}}\, f_{\rm{FD}}(r,v,v')\, \left( r^{2} \, v \, v' \right)^{\alpha} \left( \frac{n_{E11}}{W(r)} \right)^{\delta},
\end{equation}

\noindent where the coefficient $C$ (eigenvalue)  depends on the mass-loss rate $\dot{M}$ and the line-force parameter $k$ (see Eq. \ref{A5}). $W(r)$ is the dilution factor, $n_{E11}$ is the electron number density $n_{E}$ in units of $10^{-11}\, \rm{cm^{-3}}$, and $f_{\rm{FD}}$ is the finite disk correction factor. The m-CAK  line-force parameters are: $\alpha$ , $k$ and $\delta$.

The momentum equation (Eq. \ref{Smomentum}) can be expressed in a dimensionless form  \citep[see e.g.,][]{muller2008,araya2014} as:

\begin{equation}
\hat{v} \, \frac{d \hat{v}}{d \hat{r}}= -\frac{\hat{v}_{\rm{crit}}^{2}}{\hat{r}^{2}} + \hat{g}^{\rm{line}} - \frac{1}{\rho}\frac{d \rho}{d \hat{r}},
\end{equation}

\noindent with  $\hat{r}=r/R_{*}$, 
$\hat{v}=v/a$ and $\hat{v}_{\rm{crit}}=v_{\rm{esc}}/a\sqrt{2}$.
Here $R_{*}$ is the stellar radius, $a$ is the isothermal sound speed, $\hat{v}_{\rm{crit}}$ is the dimensionless rotational break-up velocity and $v_{\rm{esc}}$ is the escape velocity.
The dimensionless line acceleration reads:

\begin{equation}
\label{Snorma}
\hat{g}^{\rm{line}}=\frac{R_{*}}{a^{2}}\, g^{\rm{line}}.
\end{equation}

\noindent  Using Eq. \ref{Scontinuity} together with the equation of state \textbf{for} an ideal gas ($p=a^{2}\rho$), the dimensionless equation of motion is:

\begin{equation}
\label{Smotion}
\left( \hat{v} - \frac{1}{\hat{v}} \right) \frac{d\hat{v}}{d \hat{r}}= -\frac{\hat{v}_{\rm{crit}}^{2}}{\hat{r}^{2}} + \frac{2}{\hat{r}} + \hat{g}^{\rm{line}}.
\end{equation}

In general, the calculation of the line acceleration involves the coupling of  hydrodynamics with the radiative transport in NLTE. A very successful approach is to calculate the line acceleration using the Sobolev approximation. The pioneering work of \citet{castor1975} laid the foundations of CAK theory and later improvements (m-CAK).  A further description was done by \citet{feldmeier1998} who extended the CAK approach using a second order Sobolev approximation,  i.e,  $g^{\rm{line}}=g^{\rm{line}}(r,v,v', v'')$. \\
However, in this work, to obtain an analytical expression of the $\delta$-slow solution, we will use a radial dependence for the line acceleration following the methodology used by \citet{araya2014}, i.e., $g^{\rm{line}}=g^{\rm{line}}(r)$.  This approach allows to obtain an analytical expression for the velocity field in terms of the Lambert $W$ function (see Section \ref{SSgline}).
%
\section{Line Acceleration}
\label{SSgline}
In this section we review the basic concepts developed by \citet[][hereafter MV08]{muller2008} to derive,  later on,  a general  analytical expression for the velocity profile in the frame of the $\delta$-slow radiation-driven wind regime for massive stars. 
We demonstrate that this expression enables to integrate the equation of motion (Eq. \ref{Smotion}) leading to an analytical expression for the $\delta$-slow wind velocity profile.

\subsection{The Fast Regime Approximation}

In the framework of m-CAK stellar wind theory, MV08 present a mathematical expression for the line acceleration via a parameterized description that depends only on the radial coordinate. Using Monte Carlo multi-line radiative transfer calculations \citep{koter1997,vink1999} and a velocity profile from a $\beta$-law, these authors computed the line acceleration. Then, the numerical line acceleration, which collect all the physically motivated mathematical properties for the radiative line acceleration term,  is expressed by by the following function:

\begin{equation}
\label{SMV-gline}
\hat{g}^{\rm{line}}_{\mathrm{MV08}}(\hat{r})= \frac{\hat{g}_{0}}{\hat{r}^{1+ \delta_{1}}} \left(  1-\frac{\hat{r_{0}}}{\hat{r}^{\delta_{1}}} \right) ^{\gamma},
\end{equation}

\noindent where $\hat{g_{0}}$, $\delta_{1}$, $\hat{r_{0}}$, and $\gamma$ are the MV08 line acceleration parameters. It is important to note that these parameters, lack of any physical meaning, and besides, are not directly related to $k$, $\alpha$ and $\delta$ parameters from m-CAK theory.

Replacing Eq. \ref{SMV-gline} in Eq. \ref{Smotion}, the dimensionless equation of motion are derived and a fully analytical velocity profile is obtained (see MV08 for details about the methodology used to obtain this solution) by means of the Lambert W-function \citep{corless1993,corless1996,cranmer2004}.

The line acceleration expression given by MV08 (Eq. \ref{SMV-gline}) results in a good approximation for the m-CAK line force for $\delta \leq 0.2$, but this expression fails for $\delta$-slow solutions, when $\delta \gtrsim 0.25$. Overall, this approximation gives a poor agreement with respect to the numerical $\delta$-slow solution (from m-CAK theory). The numerical solutions are obtained from the stationary hydrodynamic code {\sc Hydwind} \citep{cure2004}.

\citet{araya2014} developed an  analytical solution for the velocity of the fast wind regime in terms of the stellar and m-CAK line-force parameters combining the methodology from MV08 and  the line acceleration proposed by \cite{villata1992}. Unfortunately,  this expression also fails when the line force parameter $\delta$ is higher than about $0.3$, because in this case a term from the proposed line acceleration expression turns complex. From a mathematical point of view,  high values of $\delta$ would require high values of $\alpha$ in order to obtain an expression with real values, but such kind of  $\alpha$ values would be totally unphysical.

\subsection{The New $\delta$-slow Regime Approximation}
\label{ourgline}

In view of the unsatisfactory results obtained when applying the approximate description of the wind velocity for the $\delta$-slow case, we decided to modify the functional form of the line acceleration given by MV08 in order to obtain a better description of the $\delta$-slow wind.  Thus, our proposed line acceleration is the following:

\begin{equation}
\label{S-gline}
\hat{g}^{\rm{line}}_{\rm{new}}(\hat{r})= \frac{\hat{g}_{0}}{\hat{r}^{1+ \delta_{1}}} \left(  1-\frac{1}{\hat{r}^{\delta_{2}}} \right) ^{\gamma},
\end{equation}

\noindent where $\hat{g}_{0}$, $\delta_{1}$, $\delta_{2}$, and $\gamma$ are the new set of line acceleration parameters. 

The new expression follows the same mathematical properties as MV08's but the inclusion of the $\delta_2$ parameter yields to a better agreement with the numerical line acceleration from the m-CAK model.

Based on this new definition for the radiation force, the new dimensionless equation of motion reads: 

\begin{equation}
\label{new-motion}
\left( \hat{v} - \frac{1}{\hat{v}} \right) \frac{d\hat{v}}{d \hat{r}}=
-\frac{\hat{v}_{{\rm crit}}^{2}}{\hat{r}^{2}} + \frac{2}{\hat{r}} 
+ \frac{\hat{g}_{0}}{\hat{r}^{1+ \delta_{1}}} \left(  1-\frac{1}{\hat{r}^{\delta_{2}}} \right) ^{\gamma}.
\end{equation}

\noindent The same methodology developed by MV08 is employed to solve the new equation of motion and the solution is given through the Lambert $W$ function,
 
\begin{equation}
\label{sol-S-MV}
\hat{v}(\hat{r})= \sqrt{-W_{j}(x(\hat{r}))},
\end{equation}

\noindent with

\begin{eqnarray}
\nonumber
x(\hat{r})& = & -\left(\frac{\hat{r}_{\rm c}}{\hat{r}} \right)^{4} \, \exp\left[-  2 \,  \hat{v}^{2}_{\rm{crit}} \left( \frac{1}{\hat{r}} - \frac{1}{\hat{r}_{\rm c}}  \right) \right. \\
& &
\left. - 2 \left(  I_{\hat{g}^{\rm{line}}}(\hat{r}) -  I_{\hat{g}^{\rm{line}}}(\hat{r}_{\rm c})   \right)  - 1 \right],
\end{eqnarray}

\noindent where 

\begin{eqnarray}
I_{\hat{g}^{\rm{line}}} &\equiv& \int \hat{g}^{\rm{line}}(\hat{r}) d \hat{r} \nonumber \\
& =&-\frac{g_{0}\, \hat{r}^{-\delta_{1}} \,{_{2}F_{1}} \left[-\gamma ,\frac{\delta_{1}}{\delta_{2}},1+\frac{\delta_{1}}{\delta_{2}},\hat{r}^{-\delta_{2}} \right]}{\delta_{1}},
\end{eqnarray}

\noindent being ${_{2}F_{1}}$ the Gauss hypergeometric function.  Note that the constant of integration vanishes due to the  subtraction between the integrals at $\hat{r}$ and $\hat{r}_{\rm c}$. The critical (or sonic) point, $\hat{r}_{\rm c}$, is obtained numerically making the RHS of Eq. \ref{new-motion} equal zero.

Finally, taking into account the numerical solution from {\sc Hydwind} as reference, a good agreement is obtained with our expression (Eq. \ref{sol-S-MV}) for the velocity profile.

\section{Line Acceleration Parameters}
\label{LAP}

In \cite{araya2014} a relationship between the MV08 line-force parameters ($\hat{g_{0}}$, $\delta_{1}$, $\hat{r_{0}}$, and $\gamma$) and the stellar and m-CAK line-force parameters was given. 
This relationship is an easy-to-use and versatile method to compute the velocity profile analytically, because both stellar and m-CAK line force parameters are already available for a wide range of spectral types \citep[see,][]{abbott1982,ppk1986,lamers1999,noebauer2015,gormaz2019,lattimer2021}. 

To derive a similar relationship, now for the $\delta$-slow regime, we created a grid of m-CAK hydrodynamic models and develop that relationship  applying a multivariate multiple regression \citep[MMR][]{rencher2012,mardia1980}.

\subsection{Grid of Hydrodynamic Models}

We built a {\sc Hydwind} grid of stellar models for $\delta$-slow solutions. The grid points were selected to cover the region of the $T_{\rm{eff}}$--$\log\,g$ diagram where the B- and A-type supergiants are located. 

For each given pair of stellar parameters ($T_{\rm{eff}}$, $\log\,g$), the stellar radius was calculated from $M_{\rm{bol}}$ by means of the flux-weighted gravity-luminosity relationship \citep{kudritzki2003,kudritzki2008}, but in addition we added 20 values for stellar radius (from 5 $R_{\sun}$ to 100 $R_{\sun}$ in steps of 5 $R_{\sun}$).  The surface gravities comprise the range of $\log g = 2.7$ down to about $90\%$ of the Eddington limit, in steps of 0.15 dex. We considered 22 effective temperature grid points, ranging from $9\,000$ K to $19\,500$ K, in steps of 500 K. These $T_{\rm{eff}}$  and $\log\,g$ ranges were adopted to describe mainly the wind of intermediate and late B supergiants.

\begin{figure}
\center
\includegraphics[width=\columnwidth]{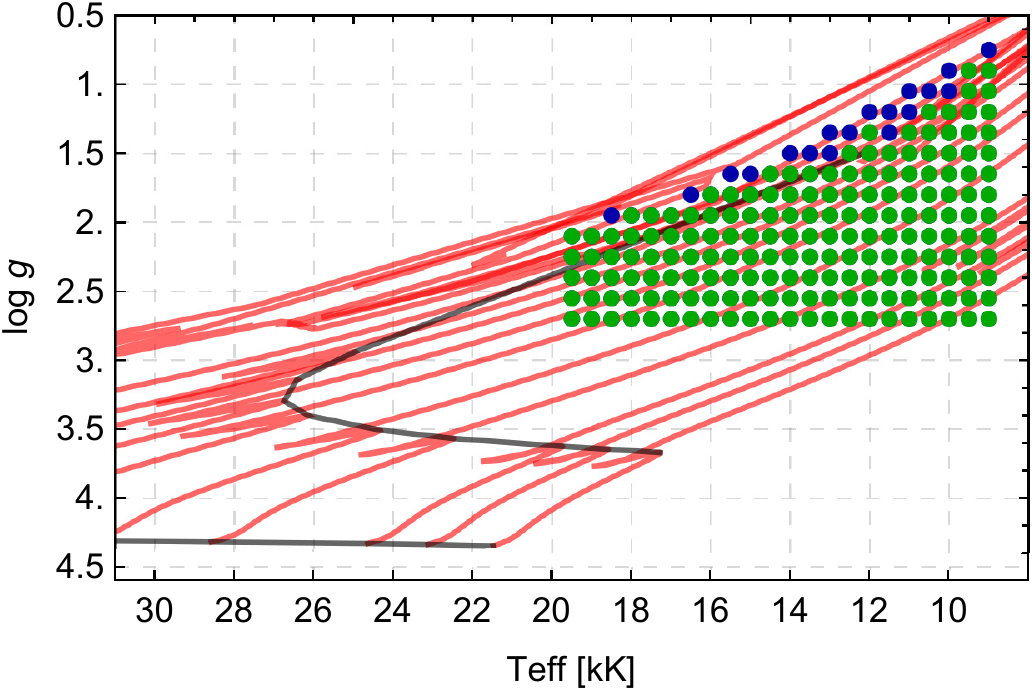}
\caption{Location of the grid models in the  $T_{\rm{eff}}$--$\log\,g$ plane. Blue and green dots represent the non-converged and converged solutions, respectively. Red solid lines represent the evolutionary tracks for stars of $7 M_{\sun}$ to $60 M_{\sun}$ without rotation \citep{ekstrom2008}, while the black lines correspond to the zero age main-sequence (ZAMS) and the terminal age main-sequence (TAMS).
\label{Dgrid}}
\end{figure}	

The m-CAK line-force parameters used for each set of $(T_{\rm{eff}}$, $\log\,g$) values are given in Table \ref{Stable1}. We considered only high values of $\delta$ in order to obtain $\delta$-slow solutions.

Then, a huge combinations of parameters were executed in {\sc Hydwind}, considering the standard boundary condition at the stellar surface, for the optical depth, $\tau_{*}=2/3$. In addition, it is worth noting that only some combinations of all parameters used in  {\sc Hydwind} converged to a physical stationary solution, i.e., we obtained $141\,067$ $\delta$-slow solutions from our initial set (about a 2\% of our initial input). In the $T_{\rm{eff}}$--$\log\,g$ plane, see Fig. \ref{Dgrid}, we show in green dots all converged models, whereas blue dots indicates that no $\delta$-slow solution was achieved for the given combination of parameters. Furthermore,  the number of converged $\delta$-slow solutions, in the  $T_{\rm{eff}}$--$\log\,g$ plane,  shows that most of the models are concentrated in the region of $\log\,g \ge 1.65$, with a peak around  $T_{\rm{eff}}=14$ kK and $\log\,g=2.4$.  Also, few models converged with values of $\delta \le 0.28$ and $\alpha \ge 0.57$. This behavior must be considered at the moment to define the limits of our approximation for $\delta$-slow solutions.

Finally, for each hydrodynamic model we fitted (Least Squares) the m-CAK line acceleration ($\hat{g}^{\rm{line}}$) with our proposed line acceleration expression (Eq. \ref{S-gline}) in order to obtain the corresponding new line acceleration parameters ($\hat{g}_{0}$, $\delta_{1}$, $\delta_{2}$, and $\gamma$).

\begin{table}
\center
\caption{Ranges of the m-CAK line-force parameters for the grid of models. \label{Stable1}}
\begin{tabular}{lcccccc}
\hline
\hline
Parameter & Range\\
\hline
$\alpha$ & 0.45 -- 0.69 (step size of 0.02) \\
$k$        & 0.05 -- 1.00 (step size of 0.05) \\
$\delta$ & 0.26 -- 0.35 (step size of 0.01)\\ 
\hline
\end{tabular}
\end{table}

\subsection{Multivariate Multiple Regression }
To derive the relationship for the new line acceleration parameters ($\hat{g}_0$, $\delta_1$, $\delta_2$ and $\gamma$) as function of stellar ($T_{\mathrm{eff}}$, $\log\mathrm{g}$, $R_{*}/R_{\sun}$) and m-CAK line-force parameters ($k$, $\alpha$, $\delta$) a MMR is applied to our grid of models.

A multiple multivariate regression model is:
\begin{equation}
Y=X B+ Z
\end{equation}

\noindent where $Y$ is a $n \times p$ matrix of data in the $p$ dependent variables, $X$ is a $n \times (1+q)$ matrix of regression: a first column of $1$'s and in the remaining columns the data of the $q$ independent variables, $B$ is a $(1+q)\times p$ matrix of parameters (the intercept and $q$ parameters, one for each of the $q$ independent variables), and $Z$ is a $n\times p$ matrix of  measurement error.

The model is the same for each dependent variable ($y_i$, $i$=$1,\dots,p$), but with different coefficients ($\beta_{i j}$, $i$=$0,\dots,p$; $j$=$0,\dots,q$), i.e.,
\begin{eqnarray}
	y_i& =& \beta_{i0} +\beta_{i1}\,T_{\mathrm{eff}}+\beta_{i2} \, \log\mathrm{g}+\beta_{i3} \,R_{*}/R_{\sun}+ \\
	& & \beta_{i4} \, k + \beta_{i5} \, \alpha+\beta_{i6} \, \delta +z \hspace{1cm}{\rm for}\,\,\, i=1, \dots, p \nonumber 
\end{eqnarray}
\noindent where $z$ represents the measurement errors. Each row of $Y$ represents an observation of each of the $p$ measured response variable.  Additional assumptions in the model are that the expectation of $Y$ is given by $E(Y)=XB$ or $E(Z)=0$, and the covariance matrix of the vectors in the rows of $Y$ is $\Sigma$, that is, the columns in $Y$ can be correlated. Also, there is an assumption of normality about the response variables that allows to perform the hypothesis testing in regression.

For our problem, the dependent variables are $\hat{g}_0$, $\delta_1$, $\delta_2$ and $\gamma$, and the independent variables are $T_{\mathrm{eff}}$, $\log\mathrm{g}$, $R_{*}/R_{\sun}$, $k$, $\alpha$ and $\delta$. The database has $n=141\,067$ records.

A data transformation is necessary to obtain a good fit of the linear model. Thus, a Box-Cox transformation \citep{seber2012} is applied to each dependent variable.  This application is performed with the public domain software  \cite{r-project}. The transformations are: $\hat{g}_0\rightarrow \hat{g}_0^{0.27}$, $\delta_1\rightarrow (\delta_1+1)^{5.3}$, $\delta_2\rightarrow\delta_2^{0.45}$, $\gamma\rightarrow (\gamma + 1)^{-3.56}$.

Finally, the estimated parameters are:

\begin{eqnarray}\label{mod:g_0}
\hat{g}_0^{0.27} &=& -4.548-  1.890 \times 10^{-4}  \, T_{\mathrm{eff}} +   \\
& & 4.393 \, \log\mathrm{g}+ 3.026 \times 10^{-2}  R_{*}/R_{\sun} -  \nonumber \\
& & 4.802 \times 10^{-3} \,  k + 3.781   \, \alpha - 3.212   \, \delta, \nonumber
\end{eqnarray}

\begin{eqnarray}\label{mod:delta1}
(\delta_1+1)^{5.3} &=& -4.623-  3.743 \times 10^{-4}  \, T_{\mathrm{eff}} +   \\
& & 1.489 \times 10^{1} \, \log\mathrm{g}+ 1.148 \times 10^{-1}  R_{*}/R_{\sun} +  \nonumber \\
& & 2.415 \,  k + 9.553 \times 10^{1}   \, \alpha - 1.320 \times 10^{2}   \, \delta, \nonumber
\end{eqnarray}

\begin{eqnarray}\label{mod:delta2}
\delta_2^{0.45} &=& 5.359 +  8.262 \times 10^{-5}  \, T_{\mathrm{eff}} -   \\
& & 1.327 \, \log\mathrm{g} - 8.327 \times 10^{-3}  R_{*}/R_{\sun} +  \nonumber \\
& & 2.181 \times 10^{-1} \,  k + 9.618 \times 10^{-1}  \, \alpha - 2.296   \, \delta \nonumber
\end{eqnarray}

\noindent and

\begin{eqnarray}\label{mod:gamma}
(\gamma + 1)^{-3.56} &=& -1.031 +  7.254 \times 10^{-6}  \, T_{\mathrm{eff}} +   \\
& & 2.994 \times 10^{-1} \, \log\mathrm{g}+ 3.097 \times 10^{-3}  R_{*}/R_{\sun} +  \nonumber \\
& & 1.836 \times 10^{-1} \,  k - 4.828 \times 10^{-1}  \, \alpha + 1.254   \, \delta, \nonumber 
\end{eqnarray}

\noindent with $R^2$ values (proportion of variability of the dependent variable explained by the regression) given in Table  \ref{Stable2}. Therefore, the regression explains almost all the variability of $\hat{g}_0^{0.27}$, a large amount of the variability of $(\delta_1+1)^{5.3}$ and $(\gamma + 1)^{-3.56}$, and a minor proportion of $\delta_2^{0.45}$.

\begin{table}
\center
\caption{Coefficient of determination ($R^2$) of the estimated models. \label{Stable2}}
\begin{tabular}{lcccccc}
\hline
\hline
Model & $R^2$\\
\hline
$\hat{g}_0^{0.27}$ & $0.9443$  \\
$(\delta_1+1)^{5.3}$ & $0.6016$\\
$\delta_2^{0.45}$ & $0.3408$\\ 
$(\gamma + 1)^{-3.56} $ & $0.7122$\\
\hline
\end{tabular}
\end{table}

After fitting the MMR, the estimated values for each dependent variable, $\hat{g}_0^{0.27}$, $(\delta_1+1)^{5.3}$, $\delta_2^{0.45}$, $(\gamma + 1)^{-3.56}$,  are obtained  and later transformed into $\hat{g}_0$, $\delta_1$, $\delta_2$, and  $\gamma$ through their respective inverse functions.

This new relationship for the line acceleration parameters ($\hat{g}_0$, $\delta_1$, $\delta_2$ and $\gamma$) as function of stellar and m-CAK line force parameters is valid only for $\delta$-slow solutions,  specifically for values of $\delta$ between 0.29 and 0.35. Therefore, it cannot be compared or used with others parametrizations obtained using an approximation for the velocity profile  of fast solution \citep[see e.g.][]{muijres2012}.

\section{The Approximative Solution}
\label{Sresults}
Once we know the relationship (estimated model) between the line acceleration parameters  as a function of the stellar and m-CAK line-force parameters, we can use Eq. \ref{sol-S-MV} to obtain the velocity profile of the $\delta$-slow wind in terms of the Lambert W-function. 

We point out that considering the number of converged models for some values of $\alpha$ and $\delta$, we limit our approximation to values of $\alpha$ between $0.45$ and $0.55$, and values of $\delta$ between $0.29$ and $0.35$. In addition, we could expect a lower precision for values of $\log\,g$ lower than  $1.65$.

In the following of this section, we discuss the accuracy of the terminal velocities and the derivation of mass-loss rates obtained using this 
\textbf{analytical} treatment.

\subsection{Terminal Velocity}

To measure the goodness of fit of the estimated model, the terminal velocity obtained by {\sc Hydwind} is compared with our formulated  solution. 

We consider two terminal velocity vectors: $v_{\infty}^{\mathrm{H}}$ defined as the terminal velocity calculated with {\sc Hydwind} (hereafter ``true terminal velocity") and $v_{\infty}^{\mathrm{A}}$ as the terminal velocity obtained from the our solution at $\hat{r}=r/R_{*}=100$, i.e., 

\begin{equation}
\label{vinf1}
v_{\infty}^{\mathrm{A}}= a\, \hat{v}_{\infty}^{\mathrm{A}} = a\,\sqrt{-W_{-1}(x(100))}.
\end{equation}

\noindent The relative error of the estimated terminal velocity $v_{\infty}^{\mathrm{A}}$ with respect to the true terminal velocity is calculated by: 
\begin{equation}
\label{Rerror}
{\rm Relative\,Error} [\%]=100 \times \frac{|v_{\infty}^{\mathrm{H}}-v_{\infty}^{\mathrm{A}}|}{v_{\infty}^{\mathrm{H}}},
\end{equation}

\noindent We obtain that the $0.90$ quantile of the distribution of the relative error are below $21\%$,  and the $0.95$ quantile of them are below $27\%$ ($q_{0.95}=27.32$).

\subsection{Mass-loss Rate}

Although our solution is developed to obtain a wind velocity profile, we can derive a recipe to obtain a  mass-loss rate. This recipe is based on the m-CAK theory, specifically  the work of \cite{cure2004}, where the velocity profile is described by our proposed  solution (Eq. \ref{sol-S-MV}). The full procedure is explained in  Appendix \ref{mdotcure}. 

Then, similar to the procedure performed for the terminal velocity, we measure the goodness of fit of the estimated mass loss rates of the models by comparing the values (vector) calculated with {\sc Hydwind}, $\dot{M}^{\mathrm{H}}$, and the ones obtained with our solution,  $\dot{M}^{\mathrm{A}}$.  

The relative error of the estimated mass-loss rate $\dot{M}^{\mathrm{A}}$ with respect to the true mass-loss rate $\dot{M}^{\mathrm{H}}$ was calculated analogously to the velocity error (Eq. \ref{Rerror}).  In comparison with the terminal velocities, the mass-loss rates have slightly higher relative errors. We observe that most of the data are below $\sim 63\%$. The $0.90$ quantile of the distribution of the relative error is below $39\%$ and the $0.95$ quantile is about $46\%$ ($q_{0.95}=46.40$).

Finally, the recipe for the calculation of the estimated mass-loss rate as a function of stellar and m-CAK line-force parameters ($T_{\mathrm{eff}}$, $\log g$, $R_{*}/R_{\sun}$, $k$, $\alpha$ and $\delta$) is the following: 
\begin{enumerate}
\item Compute $\hat{g}_0$, $\delta_1$, $\delta_2$ and $\gamma$ from Eqs. $\ref{mod:g_0}$ through $\ref{mod:gamma}$, calculating their respective inverse functions.
\item Calculate $v(r)$ from the analytical expression given in Eq. \ref{sol-S-MV}.
\item Obtain $\dot{M}^{\mathrm{A}}$ using $v(r)$ from (ii) and its gradient in the m-CAK theory  (see Appendix \ref{mdotcure}).
\end{enumerate}

\section{Discussion and Conclusions}
\label{Sconclu}

In the frame of the $\delta$-slow wind regime, we have proposed a new approximate expression for the line force based on the MV08 methodology. This new expression is  a pure function of the radial coordinate and depends on the following parameters: $\hat{g}_0$, $\delta_1$, $\delta_2$, and  $\gamma$. With this line-force we derived  an analytical expressions for the velocity profile, terminal velocity and a recipe for mass-loss rate (based on m-CAK theory and our velocity approximation). Furthermore, after generating a grid of hydrodynamic models, we apply a multivariate multiple regression to obtain a relationship among these new line-force parameters with the stellar ($T_{\rm{eff}}$, $\log g$, and $R_{*}/R_{\sun}$) and m-CAK line-force parameters ($\alpha$, $k$, and $\delta$).

The m-CAK line force parameters should be in principle self-consistently calculated coupling the hydrodynamics with the contribution to the line-acceleration from hundreds of thousand spectral lines \citep[][and references therein]{lattimer2021,gormaz2019, awa2003}. This type of calculations has not been performed for the $\delta$-slow regime, so far. Nevertheless, based on preliminary  line-profile fittings, using the $\delta$-slow solution,  \citet{cidale-praga} found that the value of $\alpha$ is in the same range as in the fast regime, while $k$ is a factor $2$-$3$ lower.

Notwithstanding we can perform a test of our  solution  using $\alpha$ and $k$ parameters from the fast  regime. To this purpose, we consider stellar and wind parameters from the work of \citet{cure2011}, where they explore the influence of ionization changes throughout the wind in the velocity profile for theoretical  models of A-type  supergiant stars. Thus, we select the models that match our grid extension (dismissing the region where $\delta \le 0.28$ and $\alpha \ge 0.57$) in order to compare it to our expression. In addition, with purpose to test the full range of our  work, we also consider the parameters from models of \citet{venero2016}, where they perform a numerical study of hydrodynamic solutions within the $\delta$-slow domain, based on fundamental parameters of typical B supergiants stars.  

The stellar and wind parameters from the mentioned works are listed in Table \ref{Ssample}. This table also  gives the values of the mass-loss rate and terminal velocity obtained from our analytical solution together with those values calculated from hydrodynamic results ({\sc Hydwind} code). All hydrodynamic models are calculated without stellar rotation.

\begin{table*}
\center
\tabcolsep 3.0 pt
\caption{Comparison of the wind parameters obtained via the new analytical solutions ($v_{\infty}^{\mathrm{{\tiny A}}}$, $\dot{M}^{\mathrm{{\tiny A}}}$) with hydrodynamic  calculations from {\sc Hydwind}  ($v_{\infty}^{\mathrm{{\tiny H}}}$, $\dot{M}^{\mathrm{{\tiny H}}}$). The models with prefix R and T  are  from \citet{cure2011} and  \citet{venero2016}, respectively.}
\label{Ssample}
\begin{tabular}{ccccccccccc}
\hline
\hline
Model & $T_{\mathrm{eff}}$ & $\log\,g$ & $R_{*}$ & $k$ & $\alpha$ & $\delta$ & $v_{\infty}^{\mathrm{{H}}}$ & $v_{\infty}^{\mathrm{{A}}}$ & $\dot{M}^{\mathrm{{H}}}$ & $\dot{M}^{\mathrm{{A}}}$\\
 & (kK) & (dex) & ($R_{\odot}$) &&&& (km\,s$^{-1}$) & (km\,s$^{-1}$) & ($10^{-6}\, M_{\odot}$\,yr$^{-1}$) & ($10^{-6}\, M_{\odot}$\,yr$^{-1}$) \\
\hline
R01 & 11.0 & 2.0 & 70 & 0.37 & 0.49 & 0.29 &  210 & 188  & 0.0052 &  0.0048\\
R02 & 11.0 & 2.0 & 70 & 0.86 & 0.49 & 0.33 &  201 & 179 & 0.20 & 0.19 \\
R05 & 11.0 & 2.0 & 60 & 0.86 & 0.49 & 0.34 &  185  & 148 & 0.15 &  0.16\\
R07& 10.0 & 2.0 & 60 & 0.37 & 0.49 & 0.30 &  207  & 161 & 0.00051 & 0.00042 \\
R08 & 10.0 & 2.0 & 60 & 0.86 & 0.49 & 0.33 &  187  & 155 & 0.017 & 0.017 \\
R11 & 10.0 & 1.7 & 80 & 0.37 & 0.49 & 0.30 &  157  & 116 & 0.0092 & 0.0091 \\
R12 & 10.0 & 1.7 & 80 & 0.86 & 0.49 & 0.34 &  152  & 106 & 0.52 & 0.61 \\
R15 & 9.5 & 2.0 & 60 & 0.37 & 0.49 & 0.30 &  193  & 162  & 0.00015 & 0.00014\\
R16 & 9.5 & 2.0 & 60 & 0.86 & 0.49 & 0.33 &  136  & 157  & 0.0048 & 0.0047\\
R19 & 9.5 & 1.7 & 100 & 0.37 & 0.49 & 0.30 &  175  & 185  & 0.0038 & 0.0031\\
R20 & 9.5 & 1.7 & 100 & 0.86 & 0.49 & 0.34 &  168  & 178  & 0.15 & 0.11\\
R23 & 9.0 & 1.7 & 100 & 0.37 & 0.49 & 0.33 &  167  & 180  & 0.00025 & 0.00019\\
R24 & 9.0 & 1.7 & 100 & 0.86 & 0.49 & 0.33 &  171  & 179  & 0.047 & 0.037\\
\hline
T15a & 15.0 & 2.11 & 52 & 0.32 & 0.50 & 0.30 &  200  & 158  & 0.90 & 0.92\\
T15b & 15.0 & 2.11 & 52 & 0.32 & 0.50 & 0.33 &  191  & 150  & 0.84  & 0.92\\
T15c & 15.0 & 2.11 & 52 & 0.32 & 0.50 & 0.35 &  186  & 144  & 0.78 & 1.00\\
T17a & 17.0 & 2.24 & 56 & 0.34 & 0.50 & 0.30 &  236  & 202  & 6.1 & 6.2 \\
T17b & 17.0 & 2.24 & 56 & 0.34 & 0.50 & 0.33 &  225  & 192  & 7.5 & 8.0 \\
T17c & 17.0 & 2.24 & 56 & 0.34 & 0.50 & 0.35 &  220  & 186  & 9.0 & 10.0 \\
T19a & 19.0 & 2.50 & 40 & 0.32 & 0.50 & 0.30 &  270  & 233  & 2.8 & 2.7 \\
T19b & 19.0 & 2.50 & 40 & 0.32 & 0.50 & 0.33 &  257  & 222  & 3.3  & 3.3\\
T19c & 19.0 & 2.50 & 40 & 0.32 & 0.50 & 0.35 &  251  &  216 & 3.8 & 4.1\\
\hline
\end{tabular}
\end{table*}

The accuracy of our approach is reflected in the  low relative errors for the mass-loss rate and terminal velocity obtained from our  solution and the hydrodynamical code. For the terminal velocity we obtain a relative error mean and median of $15.6\%$ and $15\%$, respectively. In the case of the mass-loss rate a relative error mean and median of $10.4\%$ and $7.8\%$ are obtained, respectively. \\

The use of approximate expressions that describe closely the hydrodynamics of stellar winds give the advantage of solving the radiative transfer  problem for moving media in an easy way. In particular, this new expression might properly describe the winds of late B- and A-type supergiants, without considering a $\beta$-law with high values  ($\beta$ $\gtrsim$ $3$) that lack of any physical justification in the frame of  m-CAK fast solution. 

The new expressions for the $\delta$-slow solutions together with the previously derived  expression for the fast solutions \citep{araya2014} provide an easy-to-use procedure to calculate  m-CAK wind hydrodynamics. 

Furthermore, it is important to remark that these expressions that represent the hydrodynamics of the wind can be also applied to stellar evolution codes, where mass loss rates are necessary to estimate the evolutionary phases of a star. 

In future we plan to consider the stellar rotation into our expressions and,  in addition, compare the  synthetic line profiles calculated from wind velocity profiles using a hydrodynamic code and our solutions.

\section*{Acknowledgements}
The authors would like to thank the referee,  Achim Feldmeier,  for his constructive comments.
I.A. thanks the support from FONDECYT project 11190147. I.A., M.C. \& C.A.  are grateful with the support from FONDECYT project 1190485. A.C. \& M.C. also acknowledge support from Centro Interdiciplinario de Estudios Atmosf\'ericos y Astroestad\'istica, Universidad de Valpara\'iso. C.A., A.G.-M. \& M.C. acknowledge support from Centro de Astrof\'isica de Valpara\'iso. C.A. thanks to FONDECYT project 11190945. R.O.J.V. and L.C. acknowledge financial support from the Agencia de Promoci\'on Cient\'ifica y Tecnol\'ogica (Pr\'estamo BID PICT 2016/1971), CONICET (PIP 0177), and the Universidad Nacional de La Plata (Programa de Incentivos 11/G162 and 11/G160, respectively). This project has received funding from the European Union’s Framework Programme for Research and Innovation Horizon 2020 (2014-2020) under the Marie Sk\l{}odowska-Curie Grant Agreement No. 823734. We thanks Graeme Candlish and Omar Cuervas for allowing us to use their computer facilities for the calculation of our grid.
\section*{Data Availability}

The data underlying this article will be shared on reasonable request to the corresponding author.



\bibliographystyle{mnras}
\bibliography{cites} 


\appendix
%
\section{Calculation of the Mass-loss Rate}
\label{mdotcure}

The calculation of the mass-loss rate $\dot{M}$ is obtained trough the m-CAK theory, considering the general expression for the line force and the study of \citet{cure2004}.  From this work, we can obtain the location of the singular point and the mass-loss rate using  the singularity and regularity conditions (expressed with a set of new variables).  In our case, the variables related to velocity are obtained from our proposed solution. It is important to note that this singular point is the m-CAK one and not the critical point that can be obtained from Eq. \ref{new-motion} that corresponds to the sonic point.\\
The change of variables introduced are:

\begin{equation}
u= \frac{-R_{*}}{r}, \,\, \hat{v}=\frac{v}{a}, \,\,{\rm{and}}\,\, \hat{v}'=\frac{d\hat{v}}{du}.
\end{equation}

\noindent Considering these new variables, the equation of motion reads:

\begin{eqnarray}
\nonumber
\label{cure-mo}
F(u,\hat{v},\hat{v}') & \equiv & \left( 1-\frac{1}{\hat{v}^{2}}\right) \hat{v} \frac{d\hat{v}}{du} + A + \frac{2}{u} \\ 
& & - C' \, FC \, g(u) \, (\hat{v})^{-\delta} \left( \hat{v} \frac{d\hat{v}}{du}\right)^{\alpha}=0 ,
\end{eqnarray}

\noindent where

\begin{equation}
A=\frac{G\, M(1-\Gamma)}{a^{2}R_{*}}=\frac{v^{2}_{\mathrm{esc}}}{2 a^{2}},
\end{equation}
\begin{equation}
C'=C \left(  \frac{\dot{M} D}{2 \pi} \frac{10^{-11}}{a\,R_{*}^{2}} \right)^{\delta} (a^{2} R_{*})^{(\alpha -1)},
\end{equation}
\begin{equation}
C=\Gamma GMk\left( \frac{4\pi }{\sigma _{E\;}v_{th}\;\dot{M}}\right)
^{\alpha }\;\left( \frac{D\dot{M}}{2\pi }\right) ^{\delta }  \label{A5},
\end{equation}
\noindent and
\begin{equation}
g(u)=\left(  \frac{u^{2}}{1-\sqrt{1-u^{2}}}  \right)^{\delta}. 
\end{equation}
\noindent The constant $D$ is defined as:
\begin{equation}
D=\frac{(1+Z_{\mathrm{He}} Y_{\mathrm{He}})}{(1+4\, Y_{\mathrm{He}})} \frac{1}{m_{p}},
\end{equation}
\noindent where $m_{\mathrm{p}}$ is the mass of the proton, $Y_{\mathrm{He}}$ is the helium abundance relative to hydrogen $\left( n_{\mathrm{He}}/n_{\mathrm{H}} \right) $ and $Z_{\mathrm{He}}$ is the number of free electrons provided by helium.

To calculate the location of the singular point $u_c$, and the eigenvalue, $C'$, it is necessary to 
satisfy simultaneously, 
the singularity condition,

\begin{equation}
\label{sing}
\frac{\partial}{\partial \hat{v}'}F(u,\hat{v},\hat{v}')=0\,,
\end{equation}
\noindent and the regularity condition,
\begin{equation}
\label{reg}
\frac{d}{du}F(u,\hat{v},\hat{v}')= \frac{\partial F}{\partial u} + \frac{\partial F}{\partial \hat{v}} \hat{v}'=0.
\end{equation}
\noindent Now, utilizing the change of variables
\begin{equation}
Y=\hat{v}\,\hat{v}', \,\,{\rm{and}}\,\,\, Z=\frac{\hat{v}}{\hat{v}'},
\end{equation}
\noindent Eqs. \ref{cure-mo}, \ref{sing} and \ref{reg} are expressed, respectively, as:
\begin{eqnarray}
\label{movi}
\left(  1- \frac{1}{YZ} \right)  Y + A + \frac{2}{u} -C'\, f_{1}(u,Z)g(u)Z^{-\delta /2}Y^{\alpha - \delta /2}=0,\\
\label{sing2}
\left(  1- \frac{1}{YZ} \right)  Y - C' \, f_{2}(u,Z)g(u)Z^{-\delta /2}Y^{\alpha - \delta /2}=0\\
\label{reg2}
\left(  1 + \frac{1}{YZ} \right) Y -\frac{2Z}{u^{2}} - C' \, f_{3}(u,Z)g(u)Z^{-\delta /2}Y^{\alpha - \delta /2}=0,
\end{eqnarray}
\noindent See \citet{cure2004} for the definition of $f_{1}$, $f_{2}$ and $f_{3}$.  The set of equations \ref{movi} to \ref{reg2} are valid for all known solutions from m-CAK theory.

Variables $Y$ and $Z$ are known from our proposed solution (Eq. \ref{sol-S-MV}). Now from  Eqs. \ref{sing2} and \ref{reg2}
we can solve the singular point location, $u = u_c$. Note that  $u_{c}\gtrsim0.1$ to assure a $\delta$-slow solution \citep{cure2011}.

Finally, the mass loss rate is solved from the eigenvalue, $C'$, when the singular point is replaced in Eq. \ref{cure-mo}.

%


\bsp	
\label{lastpage}
\end{document}